# Subspace tracking: a novel measurement method to test the standard phase noise model of optical frequency combs

Darko Zibar, Holger Heebøll, Jasper Riebesehl, Michael Galili, Francesco Da Ros and Aleksandr Razumov
Department of electrical and photonics engineering, DTU Electro, Technical University of Denmark, Denmark
e-mail: dazi@dtu.dk

(Invited paper)

*Summary*—The introduction of digital signal processing (DSP) assisted coherent detection has been a cornerstone of modern fiber-optic communication systems. The ability to digitally, i.e. after analogue-to-digital converter, compensate for chromatic dispersion, polarization mode dispersion, and phase noise has rendered traditional analog feedback loops largely obsolete. While analog techniques remain prevalent for phase noise characterization of single-frequency lasers, the phase noise characterization of optical frequency combs presents a greater challenge. This complexity arises from different number of phase noise sources affecting an optical frequency comb. Here, we show how a phase noise measurement techniques method based on multi-heterodyne coherent detection and DSP-based subspace tracking can be used to identify, measure and quantify various phase noise sources associated with an optical frequency comb.

*Keywords; optical frequency comb, phase noise, DSP*

## I. INTRODUCTION

While phase noise characterization of single-frequency lasers is largely solved problem [1], with commercial measurement instruments widely available, the phase noise characterization of optical frequency combs continues to pose significant challenges for the metrology community.

One reason is that the phase-noise dynamics of an optical frequency comb are inherently complex, owing to multiple fluctuating parameters, including the carrier–envelope offset, repetition rate, chirp, amplitude and pulse-width [2]. To fully characterize the phase noise of an optical frequency comb, these underlying noise sources, their correlations and their scaling with comb-line number must be measured.

In addition, many practical applications of optical frequency combs employ spectral broadening or frequency translation via nonlinear optical processes [3]. Despite some existing work in the literature [4], the contribution of nonlinear processes to the phase noise of optical frequency combs remains an open question.

Identifying and understanding the magnitude of the various phase noise sources, their mutual correlations, and their scaling with comb-line number can enable advanced stabilization schemes and, ultimately, higher measurement precision.

Current techniques for optical frequency comb phase-noise characterization rely predominantly on line-by-line measurements [3,5]. Typically, an optical bandpass filter is used to isolate individual comb lines of interest and perform subsequent phase noise measurment. While such measurements yield the total phase noise per comb line and allow the evolution of total phase noise across the comb spectrum to be inferred, they suffer from several limitations. In particular, these methods do not permit decoupling of individual phase noise sources or evaluation of their correlations. Moreover, for densely spaced optical frequency combs (line spacing < 1 GHz), optical bandpass filtering becomes technically challenging. We have recently proposed a measurement technique based on subspace tracking and multi-heterodyne coherent detection that addresses some of the aforementioned issue [6-7]. The key advantage of our approach is the ability to estimate the time evolution of individual phase noise sources and their scaling with comb-line number. These capabilities enable a complete phase-noise characterization of the optical frequency comb under test, including the power spectral density of the individual noise sources, their scaling behavior, and their correlations [6-7].

## II. BEYOND THE STANDARD PHASE NOISE MODEL

According to the standard phase noise model, it is generally understood that the total phase noise of an $m^{th}$ comb-line, $\phi_m(t)$, can be expressed as a combination of two contributions:

$$\phi_m(t) = \phi_{cm}(t) + m\phi_{rep}(t) \quad (1)$$

where $\phi_{cm}(t)$ and $\phi_{rep}(t)$ are common mode and repetition rate phase noise contributions, respectively. However, in practice an optical frequency comb can have more than two phase noise sources. Therefore, eq. (1) needs to be modified to take this into account:

$$\phi_m(t) = \phi_{cm}(t) + m\phi_{rep}(t) + a(m)\phi_a(t) + \cdots \\ + p(m)\phi_p(t) \quad (2)$$

where $\phi_a(t)$ and $\phi_p(t)$ denote residual phase noise sources, that go beyond the standard phase noise model. Since we don't know their scaling as a function of comb-line number, we state that they are a function of *m*, i.e. *a(m)* and *p(m)*. In order to get full insights of noise of an optical frequency comb, $\phi_{cm}(t)$, $\phi_{rep}(t)$, $\phi_a(t)$, …, $\phi_p(t)$, and their corresponding scaling as a function of comb line number $a(m)$, …, $p(m)$, need to be extracted. Having access to $\phi_{cm}(t)$, $\phi_{rep}(t)$, $\phi_a(t)$, …, $\phi_p(t)$, phase noise power spectral densities can be easily computed.

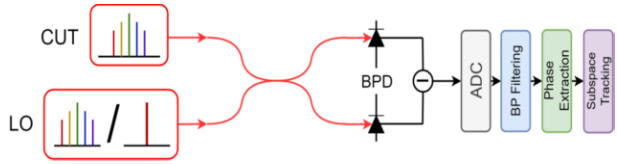

Fig. 1: Measurement technique for identification and characterization of phase noise sources affecting an optical frequency comb. CUT: comb under test, LO: local oscillator, BPD: balanced photodiode, ADC: analogue-to-digital converter, BPF: bandpass filtering.

While the state-of-the-art line-by-line measurement techniques typically measure total phase noise of a certain comb lines $\phi_m(t)$, it would be highly beneficial to have a measurement method that can measure various contribution within $\phi_m(t)$, and their corresponding scaling as a function of comb line number $m$. Such a measurement method would be able to test if an optical frequency comb follows the standard phase noise model expressed by eq. (1) or if there are additional terms beyond the standard phase noise model (as in eq. (2)).

Identifying a number of phase noise sources is important as the stabilization schemes can be designed accordingly. For instance, most of the stabilization schemes stabilize common mode and repetition rate phase noise. The presence of additional phase noise terms will then disturb the stability of the comb if not stabilized. In the following, we will describe a measurement set-up that can be used to measure phase noise sources and their scaling from eq. (2).

III. MEASUREMENT PROCEDURE FOR IDENTIFICATION OF PHASE NOISE SOURCES

In Fig. 1, a measurement set-up employing a multi-heterodyne coherent detection and subspace tracking is shown. After multi-heterodyne detection, the beat signal is digitized and digitally bandpass filtered to isolate individual comb-lines. After that phase noise estimation is employed to extract phase noise of the detected frequency components. This is used in a subspace tracking framework to find a lower-dimensional space spanned by $p$ phase noise sources. In general, we detect $M$ comb-lines resulting in $M$ phase noise traces $\phi_m(t)$, where $M = -\frac{M-1}{2} \ldots \frac{M-1}{2}$, and calculate a phase noise correlation matrix $\mathbf{C}$ of $M \times M$ dimension. Theoretically speaking, the non-zero eigenvalues of the phase noise correlation matrix $\mathbf{C}$ indicate the number of phase noise sources associated with eq. (2), while plotting the corresponding eigenvectors we obtain scaling of different noise sources as a function of comb-line number. The phase noise sources $\phi_{cm}(t)$, $\phi_{rep}(t)$, $\phi_a(t)$, …, $\phi_p(t)$, are then obtained by projecting $\phi_m(t)$ on eigenvectors representing different phase noise sources.

For example, if the correlation matrix $C$ has only two dominant eigenvalues that implies that there are only two sources of phase noise $\phi_{cm}(t)$ and $_{rep}(t)$. We have shown that this is indeed the case for an electro-optic comb, which is as expected [6]. For some optical frequency combs, we were only able to identify one dominant source of phase noise, namely $\phi_{cm}(t)$ [8-9]. The reason was that the repetition rate phase noise was below the measurement noise floor and could therefore not be measured. By applying the measurement method for phase noise characterization of a resonant electro-optic comb, we have numerically and experimentally shown that the resonant electro-optic comb does not follow the standard phase noise model at high frequencies [10]. This is due to the observation of a third phase noise component [10]. This implies that the phase noise of an $m^{th}$ comb-line, $\phi_m(t)$ of a resonant comb is described by: $\phi_{cw}(t)$, $\phi_{rep}(t)$ and $\phi_{res}(t)$. The third residual component $\phi_{res}(t)$ is a function of the $\phi_{cw}(t)$, and parameters of the resonant cavity.

We have also used subspace tracking to investigate the impact of the nonlinear spectral broadenings on the phase noise of the broadened comb. We investigated if the nonlinear processes add extra phase noise terms compared to the input comb or amplify the existing ones. We found that when a normal dispersion fiber is used the broadened combs inherits the phase noise of the input comb [11].

An important aspect of the measurement technique shown in Fig. 1(a) is the sensitivity of the phase estimation method as well as bandpass filtering of tightly spaced detected comb-lines. If the downconverted comb has narrow spacing of 10 MHz or less, it is very challenging to perform bandpass filtering. For that particular case, Kalman filtering based subspace tracking needs to be employed [7].

IV. CONCLUSIONS

Complete phase noise characterization of an optical frequency combs imposes significant challenge due to various numbers of phase noise sources affecting a comb. Measurement methods based on subspace tracking are useful for decomposition and identification of phase noise sources.

REFERENCES

[1] M. Kantner and L. Mertenskötter, "Accurate evaluation of self-heterodyne laser linewidth measurements using Wiener filters," Opt. Express 31, 15994 (2023).
[2] H. Haus and A. Mecozzi, "Noise of mode-locked lasers," IEEE J. Quantum Electron. 29, 983–996 (1993)
[3] Y. Cai et al., "On the design of low phase noise and flat spectrum optical parametric frequency comb," APL Photonics 8, 110802 (2023)
[4] A. Ruehl et al., "Ultrabroadband coherent supercontinuum frequency comb," Phys. Rev. A 84, (2011)
[5] L. Fuchuan et al. Optical linewidth of soliton microcombs. Nat Commun 13, 3161 (2022)
[6] A. Razumov et al., "Subspace tracking for phase noise source separation in frequency combs," Opt. Express 31, 34325–34347 (2023)
[7] J. Riebesehl et al., "Digital signal processing techniques for noise characterization of lasers and optical frequency combs: A tutorial," APL Photonics 9, 081101, 2024
[8] A. Razumov et al., "Phase noise characterization of a Cr:ZnS frequency comb using subspace tracking," Opt. Lett. 50, 1873-1876 (2025)
[9] A. Razumov et al., "Phase Noise Characterization of Femtosecond Laser using Subspace Tracking," in 2024 Conference on Lasers and Electro-Optics Pacific Rim (CLEO-PR), 2024, paper Mo3E_5.
[10] H. R. Heebøll et al., "Resonant EO combs: beyond the standard phase noise model of frequency combs," Opt. Express 32, 45932–45945 (2024).
[11] A. Razumov et al., "Impact of nonlinear spectral broadening on the phase noise properties of electro-optic frequency comb," https://arxiv.org/abs/2510.02868